\RequirePackage{fix-cm}
\documentclass[smallextended]{svjour3}       
\smartqed  
\usepackage{graphicx}
\usepackage{bm}
\usepackage{amsmath}
\usepackage{amsfonts}
\usepackage{amsmath}
\usepackage{amssymb}
\journalname{Few-Body Systems}
\begin{document}

\title{On formal scattering theory for differential Faddeev equations
\thanks{Supported by RFBR Grant No. 18-02-00492
}
}


\author{S. L. Yakovlev 
}


\institute{S. L. Yakovlev \at
              Saint Petersburg State University \\
              Tel.: +7 812-428-4343\\
              Fax: +7 821-428-7240\\
              \email{s.yakovlev@spbu.ru}           
              \emph{Present address:} Department of Computational Physics, Ulyanovskaya Str. 1, Saint Petersburg, 198504 Russia  
}

\date{Received: date / Accepted: date}

\maketitle

\begin{abstract}
The formal scattering theory is developed for the three-particle differential Faddeev equations. The theory is realised along the same line 
 as in the standard two-body case. The solution of the scattering problem
is  expressed in terms of the matrix T-operator constructed from the
matrix resolvent of the differential Faddeev equations.  The relationships of the matrix T-operator with elements of transition operators and Faddeev T-matrix components have been established.

\end{abstract}
\bigskip

\hfill {\it Dedicated to the memory of Ludwig Faddeev, one of my teachers.}
\section{Introduction}
\label{intro}
The progress in theoretical study of the quantum three-body problem is conventionally associated with Faddeev equations.
In its original form the Faddeev equations were formulated in \cite{Fadd60} as integral equations for T-matrix 
and wave function components for the case of short-range interactions. Later on, the Faddeev formalism  was reformulated in \cite{Lovelace64} for transition operators. The latter received the further development in \cite{AGS67} by the use of the quasi-particle method 
which reduces the Faddeev-type integral equations for the transition operators to the form of the effective multi-channel two-particle Lippmann-Schwinger (LS) equations. Another line in developing the three-body scattering formalism deals with differential form of Faddeev equations introduced in \cite{Noyes68}. This formalism was put on a sound basis in \cite{MerkGignoux76}, and its application to the three-nucleon scattering problem was demonstrated therein.
All three of these approaches are in extensive use in practical calculations of three-body collisions in atomic and nuclear systems.

In the context of the three-body Faddeev approach the three-body Faddeev differential equations (FDE) play a role of an  extension of the Schr\"odinger equation,  while the Faddeev integral equations (FIE) can be considered as a specific extension of the LS equations. 
In contrast to the two-body case there are two variants of LS equations: one corresponding to the case of free Hamiltonian resolvent taken as a driving term and the other one when the resolvent of the Hamiltonian of the system, in which only two particles interact, serves as a driving term.
FIE appeared as a remedy
for the crucial illness of the three-body LS equations: each of them do not possess the unique solutions \cite{Fadd60}. 
Faddeev integral equations
for T-matrix components \cite{Fadd60,Fadd63} are derived from the first variant of LS equation, the equations for transition operators are derived from
the set of the LS equations of the second variant.
Both variants of FIE 
formalism are equivalent in description of scattering, but the relationship between them and the DFE 
is not as straightforward as in the two-body case.

In the author's paper \cite{Yak96} the FDE 
were treated as a spectral problem for a matrix non symmetric operator. The eigenfunctions of this operator and its adjoint were constructed and the biorthogonality and completeness of eigenfunctions sets have been proven. This formed the basis of the formal scattering theory for the FDE.
It was shown that the Faddeev T-matrix components used in \cite{Fadd63} appear naturally as the matrix elements of the matrix T-operator defined from
the matrix resolvent of the operator generated by  
DFE.
Later on, some of these results  were extended on the N-particle case in \cite{Yak14} for the differential formulation of the N-particle scattering problem  
worked out  in \cite{MerkYak82-83}.
In this contribution we develop essential elements of an alternative formal scattering theory for DFE,   
where an alternative matrix T-operator  can be constructed from the transition operators.

The paper is organised as follows. Section 2 gives necessary  elements of the well  known formal scattering theory for the two-particle case introducing the key
quantities which should be generalised for the three-particle case. Section 3 deals with the three-particle scattering. Here all principal elements of the
formal scattering theory for DFE  are developed. The last section concludes the paper. Throughout the paper $\hbar=1$ is assumed.

\section{Two-particle scattering}
\label{sec:1}
The two-particle Hamiltonian $h$ in the center of mass frame with mass-weighted relative coordinates $\bm{x}\in \mathbb{R}^3$ is given by
\begin{equation}
h=h_0+v \equiv-\Delta_{\bm{x}} +v(\bm{x}),
\label{h}
\end{equation}
where $\Delta_{\bm{x}}$ is the Laplacian and $v(\bm{x})$ is a short-range potential.  The incident wave $\phi(\bm{k})$, as a solution to the equation
\begin{equation}
(h_0-k^2)\phi(\bm{k})=0,
\label{h0se}
\end{equation}
is given by the plain wave $\phi(\bm{k})=\exp\{i\bm{k}\cdot\bm{x}\}$ with $k^2=\bm{k}\cdot\bm{k}$. Here and in the following we use two notational conventions: we
systematically drop the configuration space coordinate 
in notations of the wave functions (and their components in the three-body case) when we write equations in the operator form, 
 and we use
the non bold letter for magnitude of the respective vector.

  The two-particle scattering state $\psi(\bm{k})$ is defined through the resolvent $g(z)=(h-z)^{-1}$ by the standard formula
\begin{equation}
\psi(\bm{k})=\lim_{\epsilon\to 0}(-i\epsilon)g(k^2+i\epsilon)\phi(\bm{k})
\label{psi-lim}
\end{equation}
and by construction obeys the Schr\"odinger equation
\begin{equation}
(h-k^2)\psi(\bm{k})=0.
\label{hse}
\end{equation}
The limit in (\ref{psi-lim}) can be evaluated with the help of the (second) resolvent identities  for the resolvents $g(z)$ and $g_0(z)=(h_0-z)^{-1}$. There are two of these identities
\begin{align}
g(z)=g_0(z)-g(z)vg_0(z), \label{g1} \\
g(z)=g_0(z)-g_0(z)vg_(z) \label{g2}.
\end{align}
Using  (\ref{g1}) in (\ref{psi-lim}) one obtains the representation
\begin{equation}
\psi(\bm{k})=\phi(\bm{k})-g(k^2+i0)v\phi(\bm{k}).
\label{psi-g}
\end{equation}
Using (\ref{g2}) in (\ref{psi-lim}) one ends up with the equation
\begin{equation}
\psi(\bm{k})=\phi(\bm{k})-g_0(k^2+i0)v\psi(\bm{k}),
\label{psi-ls}
\end{equation}
which is nothing but the familiar LS equation for the wave function. The representation (\ref{psi-g}) can further be transformed 
by applying the resolvent identity (\ref{g2}). This  yields
\begin{equation}
\psi(\bm{k})=\phi(\bm{k})-g_0(k^2+i0)t(k^2+i0)\phi(\bm{k}).
\label{psi-t}
\end{equation}
The operator $t(z)$ here is defined by
\begin{equation}
t(z)=v-vg(z)v
\label{t}
\end{equation}
and due to the identity
\begin{equation}
g(z)v=g_0(z)t(z)
\label{gv-g0t}
\end{equation}
satisfies the equation
\begin{equation}
t(z)=v-vg_0(z)t(z).
\label{t-ls}
\end{equation}
Another equation for $t(z)$ follows from the identity $vg(z)=t(z)g_0(z)$
\begin{equation}
t(z)=v-t(z)g_0(z)v.
\label{t-ls-1}
\end{equation}
It has been proven that the equation (\ref{t-ls}) (or equivalently (\ref{t-ls-1})) has the unique solution for a wide class of short-range 
potentials\footnote{see for instance \cite{Fadd63} and references therein} and, therefore, the formula (\ref{psi-t}) defines the unique scattering state for the Hamiltonian $h$. In its turn this scattering state provides one with the unique solution to the LS equation (\ref{psi-ls}).

The formula (\ref{psi-t}) allows us  also to determine 
the scattering amplitude. The well known asymptotic  representation for the kernel of the operator $g_0(k^2+i0)$
\begin{equation}
g_0(\bm{x},\bm{x}',k^2+i0)\sim \frac{\exp\{ikx \}}{4\pi x} \exp\{-i\bm{k}'\cdot \bm{x}'\}, \ \ x\to \infty,\  x'\ll x,
\label{g0-as}
\end{equation}
with $\bm{k}'=k\bm{x}/x$ leads in (\ref{psi-t}) to the following asymptotic representation for the scattering state as $x\to \infty$
\begin{equation}
\psi(\bm{x},\bm{k})\sim \phi(\bm{x},\bm{k}) + A(\bm{k}',\bm{k}) \frac{\exp\{ikx \}}{x}%
.
\label{psi-as}
\end{equation}
Here the scattering amplitude $A$ is given in terms of on-shell t-matrix  by
\begin{equation}
A(\bm{k}',\bm{k})= - 2\pi^2 \langle \hat{\phi}(\bm{k}')|t(k^2+i0)|\hat{\phi}(\bm{k})\rangle,
\label{A}
\end{equation}
where the normalised plane wave $\hat{\phi}(\bm{k})=(2\pi)^{-3/2}\phi(\bm{k})$ is used in the matrix element for the t-operator. Thus, the on-shell matrix elements of the t-operator defines completely the scattering amplitude and consequently the cross sections of interest. The half off-shell matrix element of the t-operator also completely defines the scattering state, as it is seen from (\ref{psi-t}), by using the standard spectral decomposition of the $g_0(z)$ resolvent
\begin{equation}
\psi(\bm{x},\bm{k})=\phi(\bm{x},\bm{k})-\int d\bm{q}\,  \frac{{\phi}(\bm{x},\bm{q})}{q^2-k^2-i0}
\langle \hat{\phi}(\bm{q})|t(k^2+i0)|\hat{\phi}(\bm{k})\rangle .
\label{psi-int}
\end{equation}
Thus, the t-operator plays the decisive role in determining of all key quantities of the scattering theory.

\section{Three-body scattering}\label{sec:3}
In this section the program of formal scattering theory exposed in the preceding section is realised for the three-particle scattering.
The three-particle Hamiltonian $H$ in the center of mass frame with mass weighted Jacobi coordinates is given by
\begin{align}
H=H_0+V,  \nonumber\\
H_0= -\Delta_{\bm{x}_\gamma}-\Delta_{\bm{y}_\gamma}, \ \ V=\sum_{\gamma=1}^3 V_{\gamma}(\bm{x}_\gamma)
,
\label{H}
\end{align}
$\bm{x}_\gamma (\bm{y}_\gamma)\in \mathbb{R}^3$.
Here and in what follows the Greek indices such as $\alpha,\beta,\gamma...$ run over the set $\{1,2,3\}$, also in the following the summation limits as in (\ref{H}) will be dropped  since the lower and the upper limits will always be 1 and 3.    The two-particle interaction potentials
$V_\gamma(\bm{x}_\gamma)$
are  assumed to be short-range as in the section \ref{sec:1}. Besides the total Hamiltonian $H$ we use the channel Hamiltonians
\begin{equation}
H_\alpha= H_0+V_\alpha.
\label{H-alpha}
\end{equation}
We use the resolvents of operators $H$, $H_\alpha$ and $H_0$ defined as
\begin{equation}
G(z)=(H-z)^{-1}, \ \ G_\alpha(z)=(H_\alpha-z)^{-1}, \ \  G_0(z)=(H_0-z)^{-1}.
\label{GGG}
\end{equation}

We consider the scattering problem, when the three-body scattering state is evolved from an eigenstate $\Phi_\beta(\bm{p}_\beta)$
of the channel Hamiltonian $H_\beta$  which is the solution to the
channel Schr\"odinger equation
\begin{equation}
(H_\beta -E) \Phi_\beta(\bm{p}_\beta)=0.
\label{H-betaSE}
\end{equation}
The explicit form of $\Phi_\beta$ reads
\begin{equation}
\Phi_\beta(\bm{X},
 \bm{p}_\beta)= \varphi_\beta(\bm{x}_\beta)\phi(\bm{y}_\beta,\bm{p}_\beta),
\label{Phi-beta}
\end{equation}
where $\bm{X}= \{ \bm{x}_\beta,\bm{y}_\beta\} \in \mathbb{R}^6$,
the two-body bound-state wave function $\varphi_\beta(\bm{x}_\beta)$ obeys the Schr\"odinger equation $(h_0+V_\beta-\varepsilon_\beta)\varphi_\beta=0$ with the binding energy $\varepsilon_\beta$. The incident momentum $\bm{p}_\beta$ is related to the energy by $E=\varepsilon_\beta +p_\beta^2$. For keeping notational simplicity we assume that for each $\beta$ there exists only one bound state of the corresponding  two-body subsystem. We will also need
the continuum wave functions $\Phi^\pm_\beta(\bm{X},\bm{p}_\beta)$ of the Hamiltonian $H_\alpha$ which are given by
\begin{equation}
\Phi^\pm_\beta(\bm{X},
\bm{k}_\beta, \bm{p}_\beta)= \psi^\pm_\beta(\bm{x}_\beta,\bm{k}_\beta)\phi(\bm{y}_\beta,\bm{p}_\beta).
\label{Phi-pm-beta}
\end{equation}
Here $\psi^\pm_{\beta}(\bm{k}_\beta)=\phi(\bm{k}_\beta)-g_{0}(k^2_\beta\pm i0)t_\beta(k^2_\beta\pm i0)\phi(\bm{k}_\beta)$
are the scattering states for particles of the pair $\beta$ defined in accordance to (\ref{psi-t}).

The scattering problem for differential Faddeev equations consists in solving the set of equations
\begin{equation}
(H_0+V_\alpha -E)\Psi_\alpha +V_\alpha\sum_{\gamma\ne \alpha} \Psi_\gamma=0, \ \ \alpha=1,2,3
\label{DFE}
\end{equation}
with $\Phi_\beta(\bm{p}_\beta)$ incident state in the $\beta$ channel and zero in all others. To proceed it is useful to introduce matrix notations. Let us define matrix operators by their matrix elements
\begin{equation}
[{\bf H}^{(0)}]_{\alpha \gamma} = (H_0+V_\alpha)\delta_{\alpha \gamma}, \ \ [\overline{\bf V}]_{\alpha \gamma}=V_\alpha(1-\delta_{\alpha \gamma}),
\label{H0V}
\end{equation}
where $\delta_{\alpha \gamma}$ is the Kronecker delta, and let us introduce vectors ${\bf \Psi}$ and ${\bf \Phi}$ with elements $ [{\bf \Psi}]_\alpha=\Psi_\alpha$ and similar for ${\bf \Phi}$.  In these notations equations
(\ref{DFE}) take the form
\begin{equation}
({\bf H}^{(0)}+\overline{\bf V}-E){\bf \Psi}=0.
\label{MDFE}
\end{equation}
The desired solution of (\ref{MDFE}) corresponding to the scattering state of interest evolved from the channel state $\Phi_\beta$ is defined with the help of the resolvent
\begin{equation}
{\bf G}(z)= ({\bf H}^{(0)}+\overline{\bf V}-z)^{-1}
\label{G}
\end{equation}
by the formula \cite{Yak96}
\begin{equation}
{\bf \Psi}^{(\beta)} =(-i\epsilon)\lim_{\epsilon \to 0} {\bf G}(E_\beta+i\epsilon){\bf \Phi}^{(\beta)},
\label{Psi-G}
\end{equation}
where $E_\beta= \varepsilon_\beta +p_\beta^2$ and the incident state vector is defined by components as $[{\bf \Phi}^{(\beta)}]_\alpha=\delta_{\alpha \beta}\Phi_\beta$.

As in the two-body case, the limit
in (\ref{Psi-G}) can be evaluated by applying the second resolvent identities for ${\bf G}(z)$ and ${\bf G}^{(0)}(z)=({\bf H}^{(0)}-z)^{-1}$.
As above, there are two variants of the second resolvent identity
\begin{align}
{\bf G}(z)={\bf G}^{(0)}(z)- {\bf G}(z)\overline{\bf V}{\bf G}^{(0)}(z), \label{G1}\\
{\bf G}(z)={\bf G}^{(0)}(z)- {\bf G}^{(0)}(z)\overline{\bf V}{\bf G}(z). \label{G2}
\end{align}
Using (\ref{G1}) for evaluation of the limit in (\ref{Psi-G}) yields the expression
\begin{equation}
{\bf \Psi}^{(\beta)}={\bf \Phi}^{(\beta)}- {\bf G}(E_\beta+i0)\overline{\bf V}{\bf \Phi}^{(\beta)}.
\label{PsiGPhi}
\end{equation}
Applying (\ref{G2}) to (\ref{Psi-G}), one arrives at the equation
\begin{equation}
{\bf \Psi}^{(\beta)}={\bf \Phi}^{(\beta)}- {\bf G}^{(0)}(E_\beta+i0)\overline{\bf V}{\bf \Psi}^{(\beta)},
\label{Psi-LS}
\end{equation}
which one immediately recognises after writing it in components
\begin{equation}
\Psi^{(\beta)}_\alpha = \delta_{\alpha \beta}\Phi_\beta - G_\alpha(E_\beta+i0)V_\alpha\sum_{\gamma\ne \alpha}\Psi^{(\beta)}_\gamma
\label{PsiIFE}
\end{equation}
as the set of FIE 
for components of the wave function \cite{Fadd60}.

The representation (\ref{PsiGPhi}) can further  be transformed by using the first of the resolvent identities (\ref{G1}). That reduces  it to the form
\begin{equation}
{\bf \Psi}^{(\beta)}={\bf \Phi}^{(\beta)}- {\bf G}^{(0)}(E_\beta+i0){\bf T}(E_\beta+i0){\bf \Phi}^{(\beta)}.
\label{PsiTPhi}
\end{equation}
Here the matrix {\bf T}-operator is defined by the 
expression
\begin{equation}
{{\bf T}}(z)= \overline{\bf V} - \overline{\bf V}{\bf G}(z)\overline{\bf V}.
\label{T}
\end{equation}
As a result, this {\bf T}-operator completely defines the solution ${\bf \Psi}^{(\beta)}$ of the scattering problem as in 
the two-body case (\ref{psi-t}).

The study of ${\bf T}(z)$ we start from deriving equations for it.
Similarly to the two-body case,  two variants of equations for ${\bf T}(z)$ 
follow from definition  (\ref{T}) if the resolvent identities (\ref{G1}) and (\ref{G2}) are
inserted into (\ref{T})
\begin{align}
 {\bf T}(z)= \overline{\bf V} - \overline{\bf V}{\bf G}^{(0)}(z){\bf T}(z), \label{T1} \\
 {\bf T}(z)= \overline{\bf V} -{\bf T}(z){\bf G}^{(0)}(z)\overline{\bf V} \label{T2}.
 \end{align}
Equations (\ref{T1}) and (\ref{T2}) are equivalent by  construction so that the solution of one satisfies  another one.
It is worth noting that the resolvent ${\bf G}(z)$ is reconstructed from the {\bf T}-operator by the standard expression
\begin{equation}
{\bf G}(z)={\bf G}^{(0)}(z)- {\bf G}^{(0)}(z){\bf T}(z){\bf G}^{(0)}(z).
\label{GT}
\end{equation}

Although equations (\ref{T1}) and (\ref{T2}) look very similar to the two-body equations (\ref{t-ls}) and (\ref{t-ls-1}), there is a significant difference,
namely the potential operator  $\overline{\bf V}$ is non symmetric whereas the two-body potential $v$ is symmetric.
This nonsymmetry leads to the different structure of source terms in (\ref{T1}) and (\ref{T2}), which is
seen from these equations written in components of matrices involved
\begin{align}
{ T}_{\alpha \beta}(z)=V_\alpha\overline{\delta}_{\alpha \beta} - V_{\alpha}\sum_{\gamma\ne\alpha}G_\gamma(z)T_{\gamma \beta}(z), \label{T11}\\
T_{\alpha \beta}(z)=V_\alpha\overline{\delta}_{\alpha \beta} -\sum_{\gamma\ne\beta}T_{\alpha \gamma}(z)G_0(z) T_{\gamma}(z). \label{T22}
\end{align}
Here $[{\bf T}]_{\alpha\beta}=T_{\alpha\beta}$,  $\overline{\delta}_{\alpha \beta}=1-{\delta}_{\alpha \beta}$  and $T_\gamma(z)=V_\gamma-V_\gamma G_\gamma(z)V_\gamma$.
While the source term of (\ref{T11}) contains the potentials, the source term of (\ref{T22}) can be written in terms of potentials or in terms of two-body
T-matrices (as it is done in (\ref{T22})) by using the identity $G_\gamma(z)V_\gamma =G_0(z)T_\gamma(z)$ (cf.   (\ref{gv-g0t})).

Equations (\ref{T11}) and (\ref{T22}) have the typical Faddeev structure of source terms and, therefore, the existence and uniqueness of their solutions are
due to the Faddeev proof given in \cite{Fadd63}. None of the equations (\ref{T11}, \ref{T22}) coincides with known equations
for T-matrix components used in \cite{Fadd63} or with
equations for transition operators used in \cite{Lovelace64} and \cite{AGS67}. In this respect (\ref{T11}, \ref{T22}) are new three-body equations of Faddeev type.
On the other hand, the equations (\ref{T22}) have identical source term with the equations for transition operators from \cite{Lovelace64,AGS67}. 
It suggests that the solution of (\ref{T22}) can be found in terms of those transition operators.
Indeed, let us write the expression
(\ref{T}) for {\bf T}-operator in matrix elements of matrices involved
\begin{equation}
T_{\alpha \beta}(z)= V_\alpha\overline{\delta}_{\alpha \beta} - \sum_{\gamma\ne\alpha}\sum_{\mu\ne\beta}V_{\alpha}G_{\gamma\mu}(z)V_{\mu}.
\label{TT}
\end{equation}
Let us now use the expression for the matrix elements 
$G_{\gamma\mu}(z)$ of the resolvent ${\bf G}(z)$ in terms of the resolvent $G(z)$ of the three-particle Hamiltonian $H$
\begin{equation}
G_{\gamma\mu}(z)=G_0(z)\delta_{\gamma\mu}(z)- G_0(z)V_\gamma G(z),
\label{GG}
\end{equation}
which can be verified by direct calculation. Inserting (\ref{GG}) into (\ref{TT}), after some algebra,
one arrives at the following representation for $T_{\alpha\beta}(z)$
\begin{equation}
T_{\alpha\beta}(z)= \overline{\delta}_{\alpha\beta}[V_\alpha+V_\alpha G_0(z)V_\alpha] - V_\alpha G_0(z)[V^\beta-V^\alpha G(z)V^{\beta}],
\label{TTT}
\end{equation}
where  $V^\alpha=\sum_{\gamma\ne\alpha}V_{\gamma}$. If one notes that
\begin{equation}
V^\beta-V^\alpha G(z)V^{\beta}= U^{(-)}_{\alpha\beta}(z)= \overline{\delta}_{\alpha\beta}(H_\alpha-z)+U_{\alpha\beta},
\label{U}
\end{equation}
where
$U^{(-)}_{\alpha\beta}(z)$ is the Lovelace transition operator \cite{Lovelace64} and $U_{\alpha\beta}$ is the AGS transition operator \cite{AGS67}, then the equations
\begin{align}
T_{\alpha\beta}(z)= \overline{\delta}_{\alpha\beta}[V_\alpha+V_\alpha G_0(z)V_\alpha] - V_\alpha G_0(z)U^{(-)}_{\alpha\beta}(z), \label{TUL} \\
T_{\alpha\beta}(z)= \overline{\delta}_{\alpha\beta}[V_\alpha+V_\alpha G_0(z)V_\alpha] - V_\alpha G_0(z)[\overline{\delta}_{\alpha\beta}(H_\alpha-z)+U_{\alpha\beta}]
\label{TUAGS}
\end{align}
establish explicit relationships between the elements of {\bf T}-operator matrix and transition operators used in \cite{Lovelace64} and \cite{AGS67}. 
For completeness we give here also the expression for $T_{\alpha\beta}(z)$ in terms of Faddeev T-matrix components \cite{Fadd63}, which are defined by
\begin{equation}
M_{\alpha\beta}(z)=V_\alpha\delta_{\alpha\beta}-V_\alpha G(z)V_\beta.
\label{M}
\end{equation}
By direct calculation one can obtain the expression
\begin{equation}
V^\beta-V^\alpha G(z)V^\beta = V_\alpha\overline{{\delta}}_{\alpha\beta} + \sum_{\mu\ne\alpha}\sum_{\nu\ne\beta}M_{\mu\nu}(z).
\label{VV}
\end{equation}
Inserting (\ref{VV}) into (\ref{TTT}), we get the relationship between our {\bf T}-operator components and  Faddeev T-matrix components
\begin{equation}
T_{\alpha\beta}(z)= \overline{\delta}_{\alpha\beta}[V_\alpha+V_\alpha G_0(z)V_\alpha] - V_\alpha G_0(z)[V_\alpha\overline{\delta}_{\alpha\beta} + \sum_{\mu\ne\alpha}\sum_{\nu\ne\beta}M_{\mu\nu}(z)].
\label{TM}
\end{equation}

The following quantity
$$
\Phi_\alpha(\bm{p}_\alpha)T_{\alpha\beta}(E_\alpha+i0),
$$
where the operator $T_{\alpha\beta}$ acts to the left, is important for application. By noting that
$$
\Phi_\alpha(\bm{p}_\alpha)= - \Phi_\alpha(\bm{p}_\alpha)V_\alpha G_0(E_\alpha+i0),
$$
one obtains 
\begin{equation}
\Phi_\alpha(\bm{p}_\alpha)T_{\alpha\beta}(E_\alpha+i0)= \Phi_\alpha(\bm{p}_\alpha)U^{(-)}_{\alpha\beta}(E_\alpha+i0)= \Phi_\alpha(\bm{p}_\alpha)U_{\alpha\beta}(E_\alpha+i0),
\label{PhiT}
\end{equation}
i.e. the left action on $\Phi_\alpha(\bm{p}_\alpha)$ of all of three operators is identical if the energy is taken properly:
$E_\alpha=\varepsilon_\alpha+p^2_\alpha$.  The same is valid also for
$\Phi^{\pm}_{\alpha}(\bm{k}_\alpha,\bm{p}_\alpha)$ if $k^2_\alpha+p^2_\alpha=E_\alpha$.

We have described in details properties of ${\bf T}(z)$ operator. Let us now turn to the solution of the scattering problem (\ref{PsiTPhi}) for wave function components.
Writing it down in components yields
\begin{equation}
\Psi^{(\beta)}_\alpha(\bm{p}_\beta)= \delta_{\alpha\beta}\Phi_\beta(\bm{p}_\beta)-G_\alpha(E_\beta+i0)T_{\alpha\beta}(E_\beta)\Phi_\beta(\bm{p}_\beta).
\label{PsiTT}
\end{equation}
By using the spectral decomposition for the resolvent $G_\alpha(z)$, we obtain the detailed representation for the components of the wave function
\begin{align}
\Psi^{(\beta)}_\alpha(\bm{X},\bm{p}_\beta)= \delta_{\alpha\beta}\varphi_\beta(\bm{x}_\beta)\phi(\bm{p}_\beta)
+\varphi_\alpha(\bm{x}_\alpha)U_{\alpha\beta}(\bm{y}_\alpha,\bm{p}_\beta,E_\beta) + U^{(0)}_{\alpha\beta}(\bm{X},\bm{p}_\beta,E_\beta),
\label{Psi-spec}
\end{align}
where $U_{\alpha\beta}$ and $U^{(0)}_{\alpha\beta}$ are given by expressions
\begin{align}
U_{\alpha\beta}(\bm{y}_\alpha,\bm{p}_\beta,E_\beta)=
- \int d\bm{p}' \frac{\phi(\bm{y}_\alpha,\bm{p}'_\alpha)}{\varepsilon_\alpha + {p'}^2_\alpha-E_\beta -i0}
\langle {\hat \Phi}_\alpha(\bm{p}'_\alpha)|T_{\alpha\beta}(E_{\beta}+i0)|{\hat \Phi}_\beta(\bm{p}_\beta)\rangle,  \label{U} \\
U^{(0)}_{\alpha\beta}(\bm{X},\bm{p}_\beta,E_\beta) = \nonumber \\
- \frac{1}{(2\pi)^{3/2}}\int d\bm{k}'_\alpha d\bm{p}'_\alpha \frac{\Phi^\pm_\alpha(\bm{X},\bm{k}'_\alpha, \bm{p}'_\alpha)}
{{k'}^2_\alpha+{p'}^2_\alpha - E_\beta -i0}
\langle {\hat \Phi^{\pm}}_\alpha(\bm{k}'_\alpha, \bm{p}'_\alpha)|T_{\alpha\beta}(E_{\beta}+i0)|{\hat \Phi}_\beta(\bm{p}_\beta)\rangle.
\label{U0}
\end{align}
Here  ${\hat \Phi}$ and ${\hat \Phi^\pm}$ are normalised states: ${\hat \Phi}=(2\pi)^{-3/2}\Phi$, ${\hat \Phi^\pm}=(2\pi)^{-3}\Phi^\pm$. As it is usual for spectral representations, the results for signs $\pm$ in (\ref{U0}) are equal to each other. From the representations (\ref{U}) and (\ref{U0}) it is possible to calculate
the asymptotic form of $U_{\alpha\beta}$ and $U^{(0)}_{\alpha\beta}$ using a technique from \cite{FaddMerk93}.
For $U_{\alpha\beta}(\bm{y}_\alpha,\bm{p}_\beta,E_\beta)$ as $y_\alpha \to \infty$, the result reads
\begin{equation}
U_{\alpha\beta}(\bm{y}_\alpha,\bm{p}_\beta,E_\beta) \sim
-2\pi^2 \langle {\hat \Phi}_\alpha(\widetilde{\bm{p}}_\alpha)|T_{\alpha\beta}(E_{\beta}+i0)|{\hat \Phi}_\beta(\bm{p}_\beta)\rangle
\frac{\exp\{ip_\alpha y_\alpha\}}{y_\alpha},
\label{U-as}
\end{equation}
where $\widetilde{\bm{p}}_\alpha=p_\alpha \bm{y}_\alpha/y_\alpha$, $p^2_\alpha=E_\beta-\varepsilon_\alpha$.
The asymptotic form of  $U^{(0)}_{\alpha\beta}(\bm{X},\bm{p}_\beta,E_\beta)$ as $X\to \infty$ is given by
\begin{align}
U^{(0)}_{\alpha\beta}(\bm{X},\bm{p}_\beta,E_\beta) \sim \nonumber \\
-2\pi^2 E^{3/4}_\beta
\langle{\hat \Phi^{-}}_\alpha(\overline{\bm{k}}_\alpha, \overline{\bm{p}}_\alpha)|T_{\alpha\beta}(E_{\beta}+i0)|
{\hat \Phi}_\beta(\bm{p}_\beta)\rangle
\frac{\exp\{i \sqrt{E_\beta} X -i\pi3/4 \}}{X^{5/2}},
\label{U0-as}
\end{align}
where $X=\sqrt{x^2_\alpha+y^2_\alpha}$,  $\overline{\bm{k}}_\alpha=\sqrt{E_\beta}\bm{x}_\alpha/X$ and $\overline{\bm{p}}_\alpha=\sqrt{E_\beta}\bm{y}_\alpha/X$.
The coefficient in front of 3D spherical wave in (\ref{U-as}) is the scattering amplitude
\begin{equation}
A_{\alpha\beta}= -2\pi^2
\langle {\hat \Phi}_\alpha(\widetilde{\bm{p}}_\alpha)|T_{\alpha\beta}(E_{\beta}+i0)|{\hat \Phi}_\beta(\bm{p}_\beta)\rangle
\label{A-ab}
\end{equation}
for elastic $\alpha=\beta$  and rearrangement $\alpha\ne \beta$ scattering.  The sum of coefficients standing in front of 6D spherical wave in (\ref{U0-as})
gives the breakup amplitude
\begin{equation}
A^{(0)}_\beta = -2\pi^2 E^{3/4}_\beta \sum_{\alpha}
\langle{\hat \Phi^{-}}_\alpha(\overline{\bm{k}}_\alpha, \overline{\bm{p}}_\alpha)|T_{\alpha\beta}(E_{\beta}+i0)|
{\hat \Phi}_\beta(\bm{p}_\beta)\rangle
,
\label{A0}
\end{equation}
since the three-body wave function is reconstructed from the components by
$$
\Psi^{(\beta)} = \sum_{\alpha}\Psi^{(\beta)}_\alpha.
$$

\section{Conclusion}
The principal elements of the formal scattering theory for Faddeev differential equations have been developed along the line which follows
the standard two-body formal scattering theory. As in the standard case, {\bf T}-operator, which is the $3\times3$ matrix, plays the central role. It is shown that all
principal quantities of the three-body scattering theory such as wave function, its Faddeev components and elastic, rearrangement, and breakup amplitudes
can be constructed from the matrix {\bf T}-operator.  
The case of scattering state evolved from the incident configuration with two bound particles and one free particle have been considered in details. 
It should be emphasized that the breakup amplitudes are calculated directly as matrix elements of {\bf T}-operator (\ref{A0}) 
between appropriate channel states similarly to elastic and rearrangement amplitudes (\ref{A-ab}). 
The scattering problem, when all of three particles are free in the incident configuration, can also
 be treated by the method developed here, and it will be done elsewhere. The derived relationship (\ref{TUAGS}) can be useful in practical applications, since 
 the well established solution method of \cite{AGS67} can be applied for calculating first $U_{\alpha\beta}$ operators and then the wave function components  
 can be calculated by applying (\ref{TUAGS}) and   (\ref{PsiTT}). 

 The inclusion of the Coulomb interaction in the formalism  is also possible in  perspective.
The modification of the FDE by splitting of long-range potentials into long-range tail part and short-range core part given in \cite{Merk80} in combination with explicit representations for wave functions of core Coulomb potential and tail Coulomb potential \cite{Yak2010} will be a basis for the Coulomb case treatment.


%
%

\begin{acknowledgements}
The author would like to thank Dr. E. Yarevsky for valuable comments.
\end{acknowledgements}


\begin{thebibliography}{}
%
%
\bibitem{Fadd60} Faddeev L. D., {\it Scattering theory for a three-particle system}, Zh. Eksperim. i Teor. Fiz. 39, 1459 (1960) [English transl. : Soviet Phys.— JETP 12, 1014 (1961)]
\bibitem{Lovelace64} Lovelace C., {\it Practical Theory of Three-Particle States. I. Nonrelativistic}, Phys. Rev., {\bf 135}, B1225 (1964)
\bibitem{AGS67} Alt E. O.,  Grassberger P.,  Sandhas W., {\it Reduction of the three-particle collision problem to multi-channel two-particle Lippmann-Schwinger equations}, Nucl. Phys., B2, 167 (1967)
\bibitem{Noyes68} Noyes P., Fideldey H., {\it Calculations of Three-Nucleon Low-Energy Parameters},  In: Three-Particle Scattering in Quantum Mechanics (Proc. of the Texas A M Conf.), (Gillespie, J., Nuttall, J., eds.), 195–294. New York: Benjamin (1968).
\bibitem{MerkGignoux76} Merkuriev S. P., Gignoux C., Lavern A., {\it Three-body scattering in configuration space}  
Ann. Phys. (N.Y.), {\bf 99}, 30 (1976)

\bibitem{Fadd63} Faddeev L. D.,  {\it Mathematical Aspects of the Three-Body Problem in Quantum Mechanics}, Israel Program for Sci. Transl.,
Jerusalem, 1965 (English trans.)
\bibitem{Yak96} Yakovlev S. L., {\it Faddeev differential equations as a spectral problem for a nonsymmetric operator},
Theor. Math. Phys., {\bf 107}, 835-847 (1996)
\bibitem{Yak14} Yakovlev S. L., {\it Quantum N-Body Problem: matrix structures and equations}, Theor. Math. Phys., {\bf 181}, 1317-1338 (2014)

\bibitem{MerkYak82-83} Merkur'ev S. P., Yakovlev S. L., {\it Differential formulation of scattering problem for N bodies},
Dans. Akad. Nauk (SSSR), {\bf 262}, 591 (1982); Merkur'ev, S. P., Yakovlev, S. L., {\it Quantum N-body scattering theory in configuration space}, 
Theor. Math. Phys., {\bf 56}, 673 (1983)
\bibitem{FaddMerk93} Faddeev L. D., Merkuriev S. P., {\it Quantum Scattering Theory for Several Particle Systems}, 406 , Kluwer, Dordrecht (1993)
\bibitem{Merk80} Merkuriev S. P., {\it On the three-body Coulomb scattering problem}, Ann. Phys., {\bf 130}, 395 (1980)
\bibitem{Yak2010} Yakovlev S. L., Volkov M. V., Yarevsky E. and Elander N.,
{\it The Impact of Sharp Screening on the Coulomb Scattering Problem in Three Dimensions},
 J. Phys. A: Math. Theor., {\bf 43}, 245302 (2010).


\end{thebibliography}


\end{document}